\documentclass{styles/svproc}
\usepackage{url}

\usepackage{graphicx} 
\usepackage{epstopdf} 
\usepackage{dcolumn} 
\usepackage{xcolor} 
\usepackage{amssymb} 
\usepackage{hyperref} 
\usepackage[utf8]{inputenc} 
\usepackage[english]{babel} 
\usepackage[displaymath, mathlines]{lineno} 
\usepackage{blindtext} 
\usepackage{physics}

\graphicspath{{figures/}} 

\def\belletwo {Belle~II}

\def\en         {\ensuremath{e^-}\xspace}   
\def\ep         {\ensuremath{e^+}\xspace}
 
\def\epem       {\ensuremath{e^+e^-}\xspace}

\def\mun        {\ensuremath{\mu^-}\xspace} 

\def\taum       {\ensuremath{\tau^-}\xspace}

\def\ellp       {\ensuremath{\ell^+}\xspace}

\def\nub        {\ensuremath{\overline{\nu}}\xspace}

\def\nub        {\ensuremath{\overline{\nu}}\xspace}

\def\nue        {\ensuremath{\nu_e}\xspace}

\def\nul        {\ensuremath{\nu_\ell}\xspace}

\def\qqbar {\ensuremath{q\overline q}\xspace}

\def\piz   {\ensuremath{\pi^0}\xspace}

\def\pip   {\ensuremath{\pi^+}\xspace}
\def\pim   {\ensuremath{\pi^-}\xspace}

\def\Kbar  {\kern 0.2em\overline{\kern -0.2em K}{}\xspace}

\def\Kz    {\ensuremath{K^0}\xspace}
\def\Kzb   {\ensuremath{\Kbar^0}\xspace}
\def\KzKzb {\ensuremath{\Kz \kern -0.16em \Kzb}\xspace}
\def\Kp    {\ensuremath{K^+}\xspace}
\def\Km    {\ensuremath{K^-}\xspace}

\def\KpKm  {\ensuremath{\Kp \kern -0.16em \Km}\xspace}
\def\KS    {\ensuremath{K^0_{\scriptscriptstyle S}}\xspace} 
\def\KL    {\ensuremath{K^0_{\scriptscriptstyle L}}\xspace}

\def\Dbar    {\kern 0.2em\overline{\kern -0.2em D}{}\xspace}

\def\Dz      {\ensuremath{D^0}\xspace}
\def\Dzb     {\ensuremath{\Dbar^0}\xspace}
\def\DzDzb   {\ensuremath{\Dz {\kern -0.16em \Dzb}}\xspace}
\def\Dp      {\ensuremath{D^+}\xspace}
\def\Dm      {\ensuremath{D^-}\xspace}

\def\DpDm    {\ensuremath{\Dp {\kern -0.16em \Dm}}\xspace}

\def\Dstarp  {\ensuremath{D^{*+}}\xspace}

\def\Bbar    {\kern 0.18em\overline{\kern -0.18em B}{}\xspace}

\def\BB      {\ensuremath{B\Bbar}\xspace} 
\def\Bz      {\ensuremath{B^0}\xspace}
\def\Bzb     {\ensuremath{\Bbar^0}\xspace}
\def\BzBzb   {\ensuremath{\Bz {\kern -0.16em \Bzb}}\xspace}
\def\Bu      {\ensuremath{B^+}\xspace}
\def\Bub     {\ensuremath{B^-}\xspace}
\def\Bp      {\ensuremath{\Bu}\xspace}

\def\BpBm    {\ensuremath{\Bu {\kern -0.16em \Bub}}\xspace}

\mathchardef\Upsilon="7107
\def\Y#1S{\ensuremath{\Upsilon{(#1S)}}\xspace}

\def\FourS {\Y4S}

\mathchardef\Deltares="7101
\mathchardef\Xi="7104
\mathchardef\Lambda="7103
\mathchardef\Sigma="7106
\mathchardef\Omega="710A

\def\Deltabar{\kern 0.25em\overline{\kern -0.25em \Deltares}{}\xspace}
\def\Lbar{\kern 0.2em\overline{\kern -0.2em\Lambda\kern 0.05em}\kern-0.05em{}\xspace}
\def\Sigbar{\kern 0.2em\overline{\kern -0.2em \Sigma}{}\xspace}
\def\Xibar{\kern 0.2em\overline{\kern -0.2em \Xi}{}\xspace}
\def\Obar{\kern 0.2em\overline{\kern -0.2em \Omega}{}\xspace}
\def\Nbar{\kern 0.2em\overline{\kern -0.2em N}{}\xspace}
\def\Xb{\kern 0.2em\overline{\kern -0.2em X}{}\xspace}

\newcommand{\tev}{\ensuremath{\mathrm{\,Te\kern -0.1em V}}\xspace}
\newcommand{\gev}{\ensuremath{\mathrm{\,Ge\kern -0.1em V}}\xspace}
\newcommand{\mev}{\ensuremath{\mathrm{\,Me\kern -0.1em V}}\xspace}
\newcommand{\kev}{\ensuremath{\mathrm{\,ke\kern -0.1em V}}\xspace}
\newcommand{\gevc}{\ensuremath{{\mathrm{\,Ge\kern -0.1em V\!/}c}}\xspace}
\newcommand{\mevc}{\ensuremath{{\mathrm{\,Me\kern -0.1em V\!/}c}}\xspace}
\newcommand{\gevcc}{\ensuremath{{\mathrm{\,Ge\kern -0.1em V\!/}c^2}}\xspace}
\newcommand{\mevcc}{\ensuremath{{\mathrm{\,Me\kern -0.1em V\!/}c^2}}\xspace}

\def\invfb   {\ensuremath{\mbox{\,fb}^{-1}}\xspace}
\def\invab   {\ensuremath{\mbox{\,ab}^{-1}}\xspace}

\def\mus  {\ensuremath{\rm \,\mus}\xspace}

\def\fs   {\ensuremath{\rm \,fs}\xspace}

\def\mus        {\ensuremath{\,\mu{\rm s}}\xspace}    

\def\to                 {\ensuremath{\rightarrow}\xspace}

\def\gsim{{~\raise.15em\hbox{$>$}\kern-.85em
          \lower.35em\hbox{$\sim$}~}\xspace}
\def\lsim{{~\raise.15em\hbox{$<$}\kern-.85em
          \lower.35em\hbox{$\sim$}~}\xspace}

\newcommand{\instPisaINFN}{INFN Sezione di Pisa, I-56127 Pisa, Italy}
\newcommand{\instPisaUNIV}{Dipartimento di Fisica, Universit\`{a} di Pisa, I-56127 Pisa, Italy}

\begin{document}
\mainmatter              
\title{Recent results from Belle and Belle II}
\titlerunning{Belle and Belle II Results}  
%
\author{Francesco Tenchini\inst{1,2} for the Belle and Belle II collaborations}
\authorrunning{Francesco Tenchini} 
%
\tocauthor{Francesco Tenchini}
\institute{\instPisaINFN
\and
\instPisaUNIV}

\maketitle              

\begin{abstract}
B-factories provide a unique environment to test the standard model in search for new physics. The Belle and \belletwo{} experiments succeeded each other at KEK, in Tsukuba, Japan. During its lifetime, Belle collected 1\invab of data; at the time of this contribution, \belletwo{} recorded 424\invfb more aiming for an unprecedented sample of 50\invab. This unique data set will allow them to search for new physics with unmatched precision. We discuss select recent results from both experiments.
\end{abstract}
\section{Introduction}
B-factories~\cite{Bevan:2014iga} are asymmetric $e^+e^-$ collider experiments designed to operate at or near the \FourS resonance energy of 10.58~\gev to produce large amounts of B mesons, D mesons and $\tau$ lepton. This allows to measure the fundamental parameters of the standard model with high precision, as well as to search for deviations from its predictions which could provide a clue of new physics. 

The Belle experiment was a general-purpose magnetic spectrometer located at the interaction region of one of such B-factories: the KEKB accelerator at KEK, in Tsukuba, Japan. Belle was composed of a silicon vertex detector, a 50-layer central drift chamber, an array of aerogel Cherenkov counters, a time-of-flight scintillation counter and an electromagnetic calorimeter. These detectors were located within a 1.5 T magnetic field provided by a superconducting solenoid magnet. Resistive plate chambers located outside the magnet detected \KL mesons and muons. Belle is described in detail in~\cite{Belle:2000cnh}. During its operation from 1999 to 2010 it recorded over 1~\invab of data, of which 711~\invfb at the $\Upsilon(4S)$ resonance corresponding to approximately 772 million \BB pairs.

Its successor \belletwo{} is located at the same facility and paired to the {SuperKEKB} collider which, following the success of KEKB, was designed to reach an unprecedented instantaneous luminosity of $6\times10^{35}$~cm$^{-2}$s$^{-1}$ through the use of a novel nano-beam collision scheme~\cite{Abe:2010gxa}. The new and harsher environment in which \belletwo{} operates demands substantial upgrades of all its subdetectors. In particular, a new tracking system was devised, composed of a two-layer silicon-pixel detector, a four-layer double-sided silicon-strip detector and a larger drift chamber, in order to provide more accurate decay vertex reconstruction closer to the interaction point.

Over its lifetime \belletwo{} aims to record 50\invab of data, enabling an extensive physics program~\cite{Belle-II:2018jsg}. At the time of this contribution \belletwo{} recorded approximately 424\invfb of collision data before entering a technical shutdown period; 190 \invfb of which were available for analysis. Although smaller than Belle's dataset, this sample can already provide significant physics input in measurements where detector performance is the dominant source of uncertainty.

In the following sections I will discuss selected recent results from both experiments. 

\section{Lifetimes of charmed particles}
Accurate prediction of the lifetime of charmed particles is challenging, as it relies on effective models such as heavy quark expansion (HQE) to calculate strong interaction contributions at low energy. Precise  measurements of these lifetimes thus provide an excellent test of these models but conversely require excellent vertex reconstruction capabilities.

The first layer of the \belletwo{} tracking system is located only 1.4~cm from the interaction point and boasts a decay-time resolution two times better than Belle, enabling high precision absolute lifetime measurements. 

\begin{figure}
    \centering
    \includegraphics[width=0.49\linewidth]{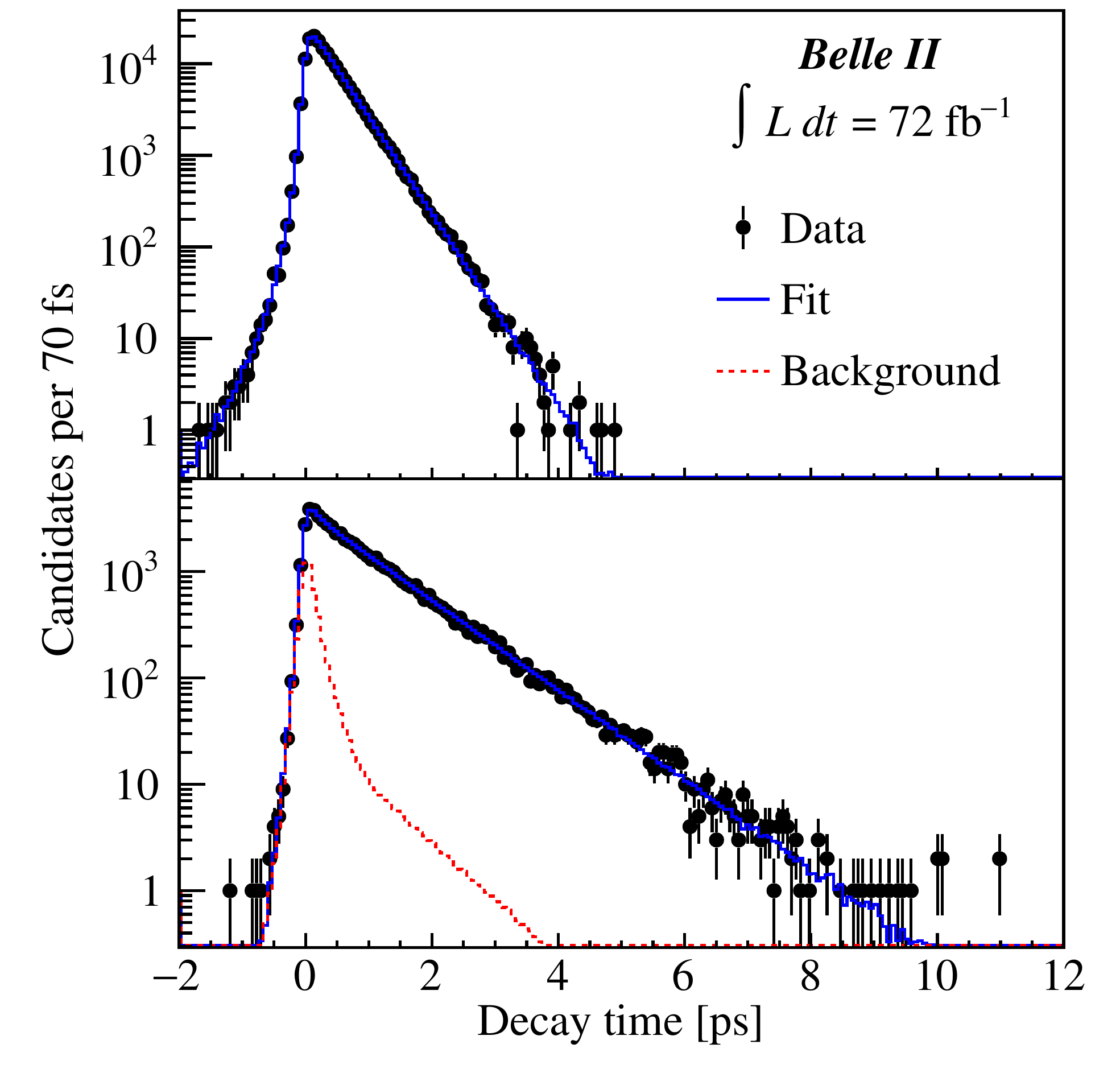}
    \includegraphics[width=0.49\linewidth]{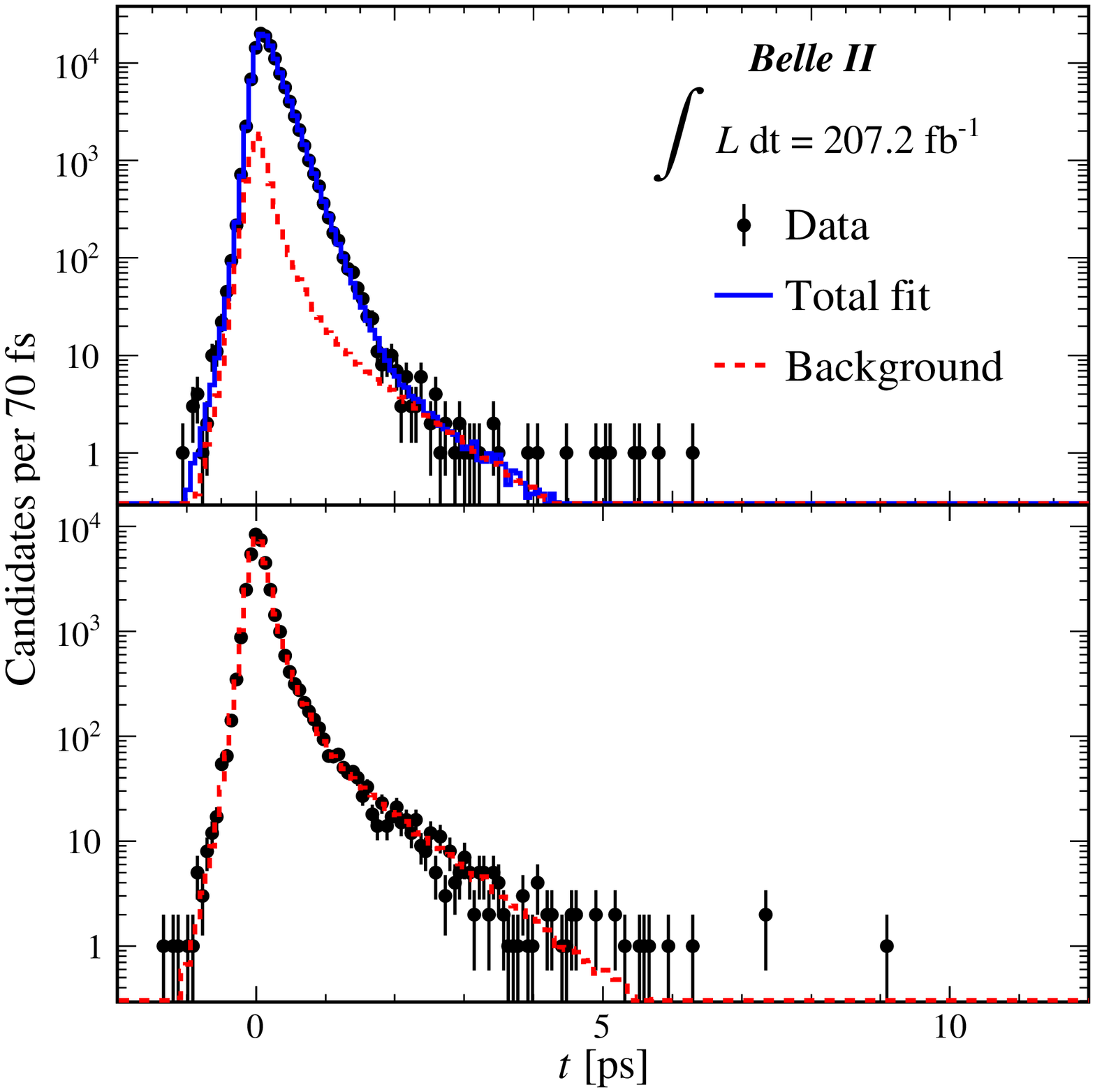}
    \caption{Decay-time distributions of $\Dz\to\Km\pip$ (\emph{top left}), $\Dp\to\Km\pip\pip$ (\emph{bottom left}) and  $\Lambda_c^+\to p K^-\pi^+$ (\emph{top right}) candidates, as well as $\Lambda_c^+\to p K^-\pi^+$ sideband events (\emph{bottom right}) with fit projections overlaid.
    }
    \label{fig:lifetimes}
\end{figure}

\belletwo{} reported measurements of the \Dz and \Dp lifetimes with an integrated luminosity of 72\invfb\cite{Belle-II:2021cxx}, and more recently of the $\Lambda_c^+$ lifetime with an integrated luminosity of 207.2\invfb\cite{Belle-II:2022ggx}. These measurements were performed respectively in the $\Dstarp\to\Dz(\to K^-\pi^+)\pi^+$, $\Dstarp\to\Dp(\to K^-\pi^+\pi^+)\pi^0$, and $\Lambda_c^+\to p K^-\pi^+$ decay channels and follow a common strategy. The \Dstarp and $\Lambda_c^+$ candidates are reconstructed from charged tracks identified as protons, pions and kaons, and from neutral pions reconstructed from two photons. A global decay-chain vertex fit\cite{Belle-IIanalysissoftwareGroup:2019dlq} constrains each decay channel to their respective topology and the \Dstarp and $\Lambda_c^+$ to originate from the interaction point.
Lifetimes are extracted with a maximum likelihood fit to the unbinned distributions of the lifetime and its uncertainty ($t$, $\sigma_t$) (Fig.~\ref{fig:lifetimes}). The background contribution in the \Dp and $\Lambda_c^+$ decay-time fits is modeled from sidebands, while in the \Dz fit the background is not modeled and assigned a systematic uncertainty. An additional uncertainty is assigned to the $\Lambda_c^+$ measurement due to contamination from $\Xi^0_c\to\pi^-\Lambda_c^+$ and $\Xi^+_c\to\pi^0\Lambda_c^+$ decays. The resulting lifetimes are 
\begin{eqnarray*}
\tau(\Dz)&=&410.5\pm1.1(\mathrm{stat})\pm0.8(\mathrm{syst})\fs,\\
\tau(\Dp)&=&1030.4\pm4.7(\mathrm{stat})\pm3.1(\mathrm{syst})\fs\mathrm{, and}\\
\tau(\Lambda_c^+)&=&203.20\pm0.89(\mathrm{stat})\pm0.77(\mathrm{syst})\fs,
\end{eqnarray*}
which are the most precise to date and in agreement with previous results, demonstrating the vertexing capabilities of the \belletwo{} detector and establishing the potential for future time-dependent analyses.

\section{Hadronic B decays}
\subsection{Combined measurement of $\phi_3/\gamma$}  
\begin{figure}
    \centering
    \includegraphics[width=0.49\linewidth]{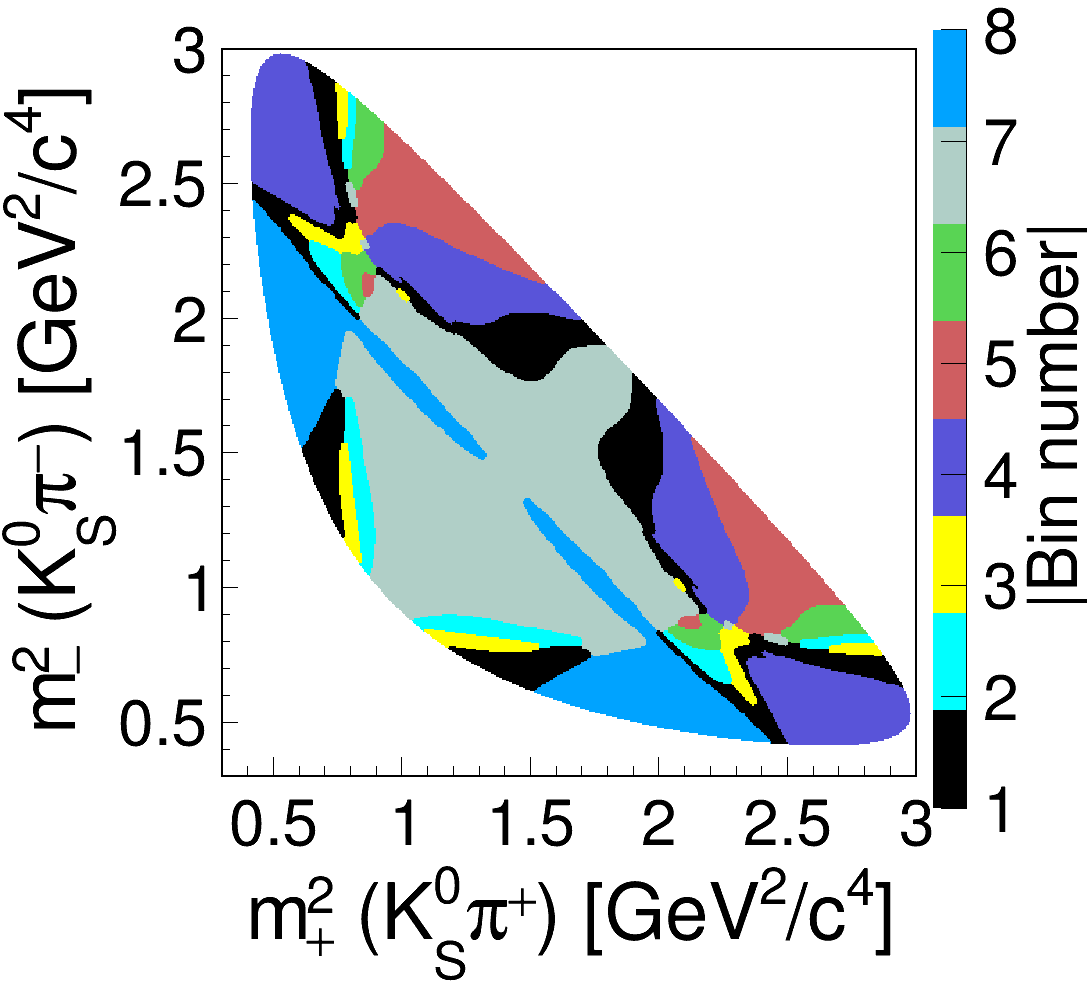}
    \includegraphics[width=0.49\linewidth]{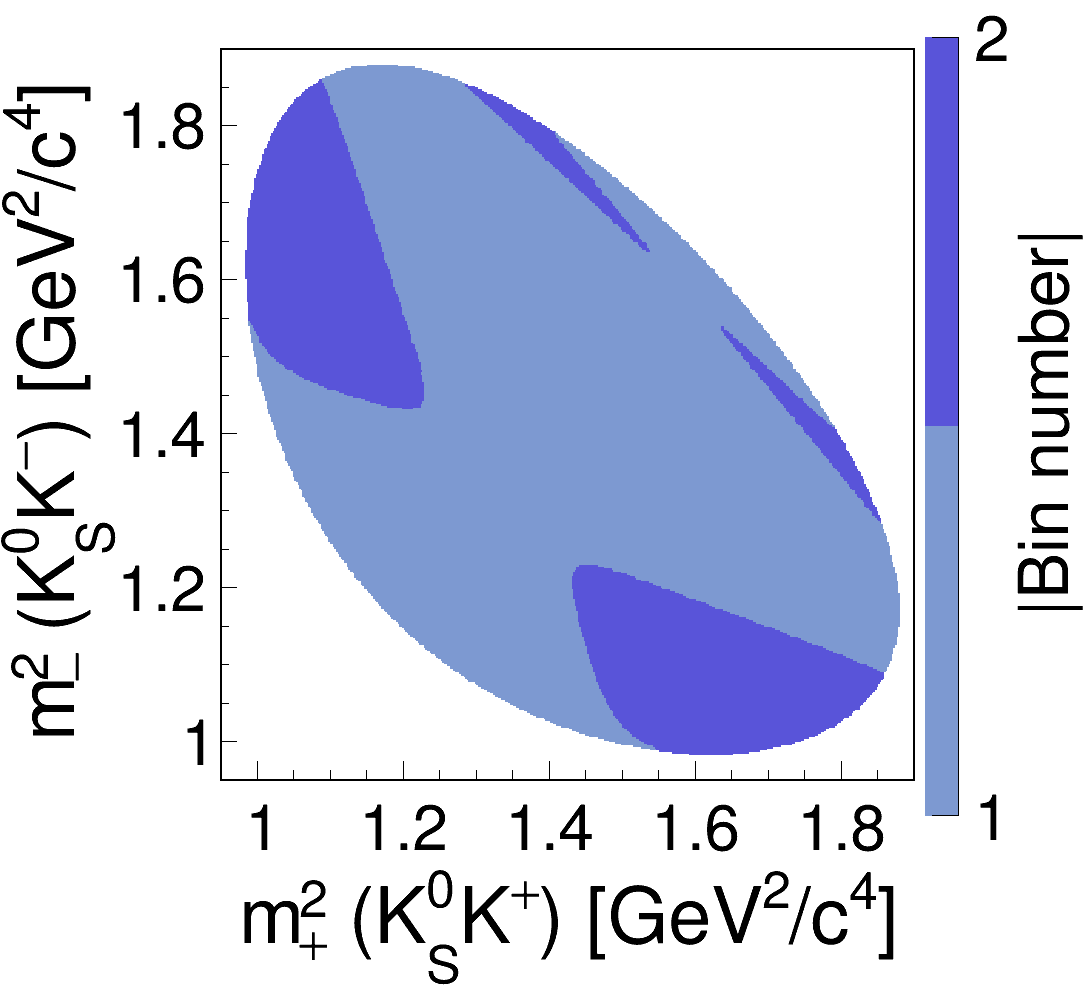}
    \caption{Binning scheme used for $\KS\pi^+\pi^-$ (\emph{left}) and  $\KS\Kp\Km$ (\emph{right}) final states.}
    \label{fig:phi3bin}
\end{figure}

The CKM angle $\phi_3/\gamma$ can be measured in a theoretically clean way from the interference of $b\to c\bar{u}s$ ($A_{fav}$) and $b\to u\bar{c}s$ ($A_{sup}$) amplitudes, the latter of which is suppressed with respect to the former. Assuming the absence of new physics at tree level, this measurement of $\phi_3$ can provide a standard model benchmark for other, indirect measurements which proceed at loop level. 
The two phases are related by
\begin{equation}
\frac{A_{sup}(B^-\to \Dzb K^-)}{A_{fav}(B^-\to \Dz K^-)} = r_B e^{i(\delta_B-\phi_3)}
\end{equation}
where $\delta_B$ is the \DzDzb strong phase. The most precise determination of $\phi_3/\gamma$ stems from a combination of LHCb measurements~\cite{Kenzie:2018oob} and has an uncertainty of $\sim 5^{\circ}$.
At B-factories, the most sensitive approach is based on the BPGGSZ method~\cite{phi3_method}, an optimally-binned (see Fig.~\ref{fig:phi3bin}) Dalitz plot analysis of $B^\pm\to D K^\pm$ and $B^\pm\to D \pi^\pm$ decays with $D$ decaying into $\KS\pi^+\pi^-$ and  $\KS\Kp\Km$ final states, where $\delta_B$ is constrained from external inputs.
Recently Belle and \belletwo{} published a combined analysis~\cite{Belle:2021efh} using 711\invfb and 128\invfb of data, respectively, which significantly improves on the previous result obtained with Belle data alone. 
The dominant background from $\epem\to\qqbar$ ($q = u,d,s,c$) events is characterised by different topology and suppressed by means of a multivariate classifier trained on shape variables. The yields in each Dalitz bin are then extracted with a simultaneously for $B^\pm\to D K^\pm$ and $B^\pm\to D \pi^\pm$ with a 2D fit in the multivariate output and in the beam-energy difference $\Delta E = \sum_j E^*_j - E^*_{beam}$, where the sum is over every signal B decay product. The result is
\begin{eqnarray*}
\phi_3 &=& (78.4 \pm 11.4 \pm 0.5 \pm 1.0)^{\circ}\\
r_B^{DK} &=& 0.129 \pm 0.024 \pm 0.001 \pm 0.002\\
\delta_B^{DK} &=& (124.8 \pm 12.9 \pm 0.5 \pm 1.7)^{\circ}
\end{eqnarray*}
where the first uncertainty is statistical, the second is systematic, and the third originates from external input on the strong phase. The uncertainty on $\phi_3$ improved from the $15^{\circ}$ (statistical) and  $4^{\circ}$ (systematic) of the previous measurement thanks to better \KS reconstruction, background suppression and analysis strategy. The uncertainty related to the strong-phase  also improved thanks to new measurements from BESIII~\cite{BESIII:2020khq,BESIII:2020hpo}. Although the uncertainty is still larger than the current world-average value, it is also statistically limited; a future \belletwo{} analysis with a data set corresponding to 10\invab could lower it to approximately $4^{\circ}$.

\subsection{Charmless decays at \belletwo} 
Charmless hadronic two-body decays $B\to K\pi$ provide an opportunity to test the standard model by studying direct CP violation which is sensitive to new physics contributions. In the standard model the isospin-sum rule~\cite{Gronau:2005kz}
\begin{equation}
\begin{aligned}
    \mathcal{A}_{CP}(\Kp\pi^-)
    +\mathcal{A}_{CP}(\Kz\pi^+)\frac{\mathcal{B}(\Kz\pi^+)}{\mathcal{B}(\Kp\pi^-)}\frac{\tau_{B^0}}{\tau_{B^+}}\\
    -2\mathcal{A}_{CP}(\Kp\pi^0)\frac{\mathcal{B}(\Kp\pi^0)}{\mathcal{B}(\Kp\pi^-)}\frac{\tau_{B^0}}{\tau_{B^+}}
    -2\mathcal{A}_{CP}(\Kz\piz)\frac{\mathcal{B}(\Kz\pi^0)}{\mathcal{B}(\Kp\pi^-)} =0,
\end{aligned}
\end{equation}
where ${A}_{CP}$ are direct CP asymmetries, holds true within 1\%. The precision of this null test is dominated by experimental uncertainties on the $B\to\KS\piz$ decay mode, which can be accurately studied only at B-factories.
Belle II reports a measurement of this process using 190\invfb of data~\cite{Belle-II:2022jij}.
B meson candidates are built using \KS and \piz candidates; the flavour of the other B is determined from the remaining particles in the event using a dedicated multivariate algorithm~\cite{Belle-II:2021zvj}. The branching fractions and $\mathcal{A}_{CP}$ are then extracted by an extended maximum-likelihood fit of the distributions of $\Delta E$, $\Delta t$, a modified version of the beam-constrained mass  $M_{bc} = \sqrt{{E^*_{beam}}^2-\vec{p}^2_B}$, and a continuum-suppressing classifier.
The results are 
\begin{eqnarray*}
\mathcal{B}(B\to\KS\piz)&=&11.0\pm1.2(\mathrm{stat})\pm1.0(\mathrm{syst})]\times10^{-6}\\
\mathcal{A}_{CP}&=&-0.41^{+0.30}_{-0.32}(\mathrm{stat})\pm0.09(\mathrm{syst})
\end{eqnarray*}
 in agreement with the current world average.

The radiative version of this decay, $B\to\KS\piz\gamma$, is also interesting as $b\to s\gamma$ transitions only occur at loop level in the standard model and with flavour-specific polarization. The photon is right-handed in \Bz decays and left-handed in \Bzb decays. Therefore, we don't expect to observe any time-dependent asymmetry. However, new physics occurring at tree level with different chirality rules could alter this picture. \belletwo{} measured the branching fraction of this decay in preparation for a time-dependent analysis. The result, {$\mathcal{B}=(7.3\pm1.8(\mathrm{stat})\pm1.0(\mathrm{syst}))\times10^{-6}$}, is once again in agreement with the world average value.

\section{Semileptonic B decays}
\subsection{Status of $|V_{ub}|$ and $|V_{cb}|$}
The CKM matrix elements $|V_{ub}|$ and $|V_{cb}|$ are measured through the study of tree-level semileptonic B decays, where the standard model contribution is assumed to dominate. Such decays can be analyzed either exclusively or inclusively; in the former case a specific decay mode is reconstructed, e.g. $B\to\pi\ell\nu$, while in the latter every final state is reconstructed irrespectively of the hadronic composition: $B\to X_u\ell\nu$. The two methods, which should be in agreeement, are instead subject to a long-standing $3\sigma$ discrepancy~\cite{HFLAV:2019otj}. B-factories are uniquely suited to probe this puzzle by tackling the measurement in both approaches.

\subsection{Partial branching fractions of $B\to X_u\ell\nul$ at Belle}
Inclusive measurements of $B\to X_u\ell\nul$ are extremely challenging due to the large background from favored $B\to X_c\ell\nul$ decays, unless the measurement is performed in regions where the latter are kinematically forbidden. However, doing so introduces large theory uncertainties on the decay rate from to nonperturbative shape functions.
In absence of as-of-yet-unrealized modeling improvements or model-independent approaches, it is best to extend the measurement into the $B\to X_c\ell\nul$ dominated region. Background suppression is therefore of critical importance. 

To perform this measurement, Belle reconstructs the second B meson in a variety of hadronically decaying modes using a multivariate algorithm~\cite{Feindt:2011mr}.
The modeling of $B\to X_c\ell\nul$ is then verified in charm-enriched sidebands; a background-suppressing classifier is trained using several discriminating observables, such as the missing squared mass or the number of observed kaons in the decay. The partial branching fraction is extracted from a template fit to the distributions of remaining events, obtaining~\cite{Belle:2021eni}
\begin{eqnarray*}
\Delta\mathcal{B}=[1.59\pm0.07(\mathrm{stat})\pm0.16(\mathrm{syst})]\times10^{-3}.
\end{eqnarray*}
This branching fraction probes about 86\% of the phase space available to the decay. It is then converted to $|V_{ub}| = (4.10\pm0.09\pm0.22\pm0.15)\times10^{-3}$, where the last uncertainty denotes the theory error, using the average of several theory predictions. This result is consistent with previous inclusive determinations~\cite{BaBar:2016rxh} and deviates from exclusive results by $1.3\sigma$.

\subsection{$B\to \pi e\nu$ at \belletwo{}}
Tagged measurements at \belletwo{} are limited by the available sample size; nevertheless, they are important to set the framework for future analyses. \belletwo{} produced a first measurement of the exclusive branching fractions of $\Bz\to\pim\ep\nue$ and $\Bp\to\piz\ep\nue$ 
on 190\invfb of data using a hadronic tagging approach similar to the previous section, but using a novel Full Event Interpretation algorithm~\cite{Keck:2018lcd} recently developed for \belletwo. The results are~\cite{Belle-II:2022pir} 
\begin{eqnarray*}
&\mathcal{B}(\Bz\to\pim\ep\nue)=(1.43\pm0.27(\mathrm{stat})\pm0.07(\mathrm{syst}))\times10^{-4}\\
&\mathcal{B}(\Bp\to\piz\ep\nue)=(8.33\pm1.67(\mathrm{stat})\pm0.55(\mathrm{syst}))\times10^{-5}\\
&|V_{ub}| = (3.88\pm0.45)\times10^{-3} 
\end{eqnarray*}
the latter of which is extracted with a combined fit using input from lattice QCD. The result agrees with pre-existing exclusive determinations.

\subsection{$|V_{cb}|$ with $q^2$ moments of $B\to X_c\ell\nu$}
The inclusive determination of $|V_{cb}|$ proceeds through the measurement of the differential decay rate of $B\to X_c\ell\nu$, which using HQE can be written as a power expansion of the inverse b-quark mass, $1/m_b$. The coefficients of this expansion can be directly measured on experimental data, typically from moments of the lepton energy and invariant hadronic mass, greatly reducing uncertainty from theory input. However, at higher orders of $1/m_b$ the number of these coefficients increases rapidly making the inclusion of higher power corrections increasingly difficult. 

Following a recent idea~\cite{Fael:2018vsp}, both Belle~\cite{Belle:2021idw} and \belletwo{}~\cite{Belle-II:2022fug} measured the moments of the dilepton moment ($q^2$) spectrum for the first time. These quantities depend on a reduced set of parameters, making a fully data driven extraction of $|V_{cb}|$ possible up to $1/m_b^4$. The results from the two experiments were subsequently used in a global fit~\cite{Bernlochner:2022ucr} to extract $|V_{cb}| = (41.69\pm0.59\pm0.23)\times 10^{-3 }$, where the first uncertainty comes from the fit and the second from higher order contributions to the power expansions. This is in good agreement with the measurement performed through the traditional approach~\cite{Bordone:2021oof} and can be further improved in the future, placing extremely strong constraints on this element of the CKM matrix.

\section{Tests of lepton flavour}
\subsection{$\Omega_c^0\to\Omega^-\ellp\nul$ at Belle}
Electroweak coupling of gauge bosons in the standard model is assumed identical for every lepton generation; this is known as lepton flavour universality (LFU). The recent emergence of anomalies 
such as in $R_K = \frac{\mathcal{B}(B\to K\mu^+\mu^-)}{\mathcal{B}(B\to Ke^+e^-)}$~\cite{LHCb:2021trn} suggest this might not be the case, which would be a clear sign of new physics.

Semileptonic decays of charmed baryons play an important role in the study of weak and strong interactions but are relatively challenging due to either low production rates or complicated backgrounds. 
The decay $\Omega_c^0\to\Omega^-\ellp\nul$ in particular had never been tested for LFU. Belle performed the measurement~\cite{Belle:2021dgc} using a combined sample of 89.5\invfb, 711\invfb and 121.1\invfb, taken at 10.52, 10.58 and 10.86\gev center-of-mass energy, respectively. The semileptonic yields are measured relatively to their hadronic counterpart, $\Omega_c^0\to\Omega^-\pip$ by means of a binned maximum likelihood fit. They are then corrected for their relative effieciencies, resulting in the branching-fraction ratios
\begin{eqnarray*}
\mathcal{B}(\Omega_c^0\to\Omega^-e^+\nu_e)/\mathcal{B}(\Omega_c^0\to\Omega^-\pi^+) &=& 1.98\pm0.13\pm0.08\\
\mathcal{B}(\Omega_c^0\to\Omega^-\mu^+\nu_\mu)/\mathcal{B}(\Omega_c^0\to\Omega^-\pi^+) &=& 1.94\pm0.18\pm0.10\\
\mathcal{B}(\Omega_c^0\to\Omega^-e^+\nu_e)/\mathcal{B}(\Omega_c^0\to\Omega^-\mu^+\nu_\mu) &=& 1.02\pm0.10\pm0.02 .
\end{eqnarray*}
where the first uncertainty is statistical and the second systematic.
The first result greatly improves on the previous maesurement~\cite{CLEO:2002imi}; the second has never been measured before; and the third is also new and in good agreement with standard-model LFU predictions.

\subsection{$\tau^\pm\to\ell^\pm\gamma$ at Belle}
A related underlying assumption of the standard model is that the flavour of charged leptons is itself conserved, with immeasurably small violations becoming possible at loop level through neutrino oscillation. The same is not true in many new physics models, which envision scenarios in which lepton flavour is violated. Models predicting flavour violation between the first and the second lepton family are tightly constrained by experimental evidence, e.g.~\cite{MEG:2016leq}; suggesting that, if indeed flavour is violated, this should occur in the third generation.

B-factories have been historically at the forefront of $\tau$ flavour violation searches, with Belle and BaBar providing most of the constraints such decays. Most recently, Belle updated their measurement of the $\tau\to\ell\gamma$ decay~\cite{Belle:2021ysv} which was previously performed on a 535\invfb data sample. An improved selection provides better suppression of the dominant background stemming from the coincidence of $\tau\to\ell\nu\nub$ decays with spurious photons. Belle places an upper limit to this branching fraction, at the 90\% confidence level, of $\mathcal{B}(\taum\to\en\gamma) <5.6\times 10^{-8}$ and $\mathcal{B}(\taum\to\mun\gamma) <4.2\times 10^{-8}$. The limit on the muon channel is the most stringent to date, superseding the previous value from BaBar~\cite{BaBar:2009hkt}.

\subsection{$B^0\to\tau^\pm\ell^\mp$ at Belle}
Flavour could also be violated in tauonic B decays through loops involving NP particles such as leptoquarks~\cite{Smirnov:2018ske}
This process is kinematically similar to $B^0\to D^{(*)-}\pi^+$, which constitutes the main experimental background. 
Rather than explicitly reconstructing the $\tau$ decays, which can prove difficult due to the presence of neutrinos, Belle instead reconstructs~\cite{Belle:btotauell} the other B (tag) and the light charged particle (a lepton in the signal case, a pion for the background). Energy conservation is then used to calculate the missing mass of the system, resulting in characteristic peaks at the mass of the D, D* and tau particles, with the addition of an exponential-like combinatorial background.

Signal yields are extracted with an unbinned extended maximum likelihood fit, and no signal evidence is found, resulting in an upper limit on the branching fraction, at 90\% confidence level, of
\begin{eqnarray*}
\mathcal{B}(B^0\to\tau^\pm e^\mp) < 1.5\times10^{-5}\\
\mathcal{B}(B^0\to\tau^\pm\mu^\mp) < 1.6\times10^{-5}
\end{eqnarray*}
The bound on the electron channel is the most stringent yet, while the one on the muon channel is similar to the pre-existing limit from LHCb~\cite{LHCb:2019ujz}.

\section{Conclusions}
B factories provide a unique environment for precision measurements. The Belle experiment, although long past its operation date, is still producing abundant physics output thanks to its large data sample. Meanwhile \belletwo{} has entered active operation and collected so far 424\invfb of data, or approximately half that of Belle; which has already been used to provide high impact measurements. Ultimately, \belletwo{} aims to collect 50 times the Belle data set, allowing it to search for new physics with unprecedented precision.

\section*{Acknowledgements}
These proceedings have received funding from the European Union’s Horizon 2020 research and innovation programme under the Marie Skłodowska-Curie grant agreement No 101026516.

%


\begin{thebibliography}{99}
%

\bibitem{Bevan:2014iga}
A.~J.~Bevan \textit{et al.} (BaBar and Belle collaborations), 
Eur. Phys. J. C 74, 3026 (2014)

\bibitem{Belle:2000cnh}
A.~Abashian \textit{et al.} (Belle collaboration),
Nucl. Instrum. Meth. A \textbf{479} (2002), 117-232

\bibitem{Abe:2010gxa}
T.~Abe \textit{et al.} (Belle II collaboration), 

\bibitem{Belle-II:2018jsg}
E.~Kou \textit{et al.} (Belle II collaboration), 
PTEP \textbf{2019} (2019) no.12, 123C01
[erratum: PTEP \textbf{2020} (2020) no.2, 029201]

\bibitem{Belle-II:2021cxx}
F.~Abudin\'en \textit{et al.} (Belle II collaboration),
Phys. Rev. Lett. \textbf{127} (2021) no.21, 211801

\bibitem{Belle-II:2022ggx}
F.~Abudin\'en \textit{et al.} (Belle II collaboration),
[arXiv:2206.15227 [hep-ex]].

\bibitem{Belle-IIanalysissoftwareGroup:2019dlq}
J.~F.~Krohn \textit{et al.} (Belle II analysis software group),
Nucl. Instrum. Meth. A \textbf{976} (2020), 164269

\bibitem{Kenzie:2018oob}
M.~W.~Kenzie \textit{et al.} (LHCb collaboration),
LHCb-CONF-2018-002

\bibitem{phi3_method}
A. Giri, Y. Grossman, A. Soffer, J. Zupan, 
Phys. Rev. D \textbf{68} (2003) 054018

\bibitem{Belle:2021efh}
F.~Abudin\'en \textit{et al.} (Belle and Belle II collaborations),
JHEP \textbf{02} (2022), 063

\bibitem{BESIII:2020khq}
M.~Ablikim \textit{et al.} (BESIII collaboration),
Phys. Rev. D \textbf{101} (2020) no.11, 112002

\bibitem{BESIII:2020hpo}
M.~Ablikim \textit{et al.} (BESIII collaboration),

\bibitem{Belle-II:2022jij}
F.~Abudin\'en \textit{et al.} (Belle II collaboration),
[arXiv:2206.07453 [hep-ex]].

\bibitem{Gronau:2005kz}
M.~Gronau,
Phys. Lett. B \textbf{627} (2005), 82-88

\bibitem{Belle-II:2021zvj}
F.~Abudin\'en \textit{et al.} (Belle II collaboration),
Eur. Phys. J. C \textbf{82} (2022) no.4, 283

\bibitem{HFLAV:2019otj}
Y.~S.~Amhis \textit{et al.} (HFLAV),
Eur. Phys. J. C \textbf{81} (2021) no.3, 226

\bibitem{Feindt:2011mr}
M.~Feindt, F.~Keller, M.~Kreps, T.~Kuhr, S.~Neubauer, D.~Zander and A.~Zupanc,
Nucl. Instrum. Meth. A \textbf{654} (2011), 432-440

\bibitem{Keck:2018lcd}
T.~Keck, F.~Abudin\'en, F.~U.~Bernlochner, R.~Cheaib, S.~Cunliffe, M.~Feindt, T.~Ferber, M.~Gelb, J.~Gemmler and P.~Goldenzweig, \textit{et al.}
Comput. Softw. Big Sci. \textbf{3} (2019) no.1, 6

\bibitem{Belle:2021eni}
L.~Cao \textit{et al.} (Belle collaboration),
Phys. Rev. D \textbf{104} (2021) no.1, 012008

\bibitem{BaBar:2016rxh}
J.~P.~Lees \textit{et al.} (BaBar collaboration),
Phys. Rev. D \textbf{95} (2017) no.7, 072001

\bibitem{Belle-II:2022pir}
F.~Abudin\'en \textit{et al.} (Belle II collaboration),
[arXiv:2206.08102 [hep-ex]].


\bibitem{Fael:2018vsp}
M.~Fael, T.~Mannel and K.~Keri Vos,
JHEP \textbf{02} (2019), 177


\bibitem{Belle:2021idw}
R.~van Tonder \textit{et al.} (Belle collaboration),
Phys. Rev. D \textbf{104} (2021) no.11, 112011

\bibitem{Belle-II:2022fug}
Belle II collaboration,
[arXiv:2205.06372 [hep-ex]]

\bibitem{Bernlochner:2022ucr}
F.~Bernlochner, M.~Fael, K.~Olschewsky, E.~Persson, R.~van Tonder, K.~K.~Vos and M.~Welsch,
JHEP \textbf{10} (2022), 068

\bibitem{Bordone:2021oof}
M.~Bordone, B.~Capdevila and P.~Gambino,
Phys. Lett. B \textbf{822} (2021), 136679

\bibitem{LHCb:2021trn}
R.~Aaij \textit{et al.} (LHCb collaboration),
Nature Phys. \textbf{18} (2022) no.3, 277-282

\bibitem{Belle:2021dgc}
Y.~B.~Li \textit{et al.} (Belle collaboration),
Phys. Rev. D \textbf{105} (2022) no.9, L091101

\bibitem{CLEO:2002imi}
R.~Ammar \textit{et al.} (CLEO collaboration),
Phys. Rev. Lett. \textbf{89} (2002), 171803

\bibitem{MEG:2016leq}
A.~M.~Baldini \textit{et al.} (MEG collaboration),
Eur. Phys. J. C \textbf{76} (2016) no.8, 434

\bibitem{Belle:2021ysv}
A.~Abdesselam \textit{et al.} (Belle collaboration),
JHEP \textbf{10} (2021), 19

\bibitem{BaBar:2009hkt}
B.~Aubert \textit{et al.} (BaBar collaboration),
Phys. Rev. Lett. \textbf{104} (2010), 021802

\bibitem{Smirnov:2018ske}
A.~D.~Smirnov,
Mod. Phys. Lett. A \textbf{33} (2018), 1850019

\bibitem{Belle:btotauell}
H.~Atmacan \textit{et al.} (Belle collaboration),
Phys. Rev. D \textbf{104} (2021) no.9, L091105

\bibitem{LHCb:2019ujz}
R.~Aaij \textit{et al.} (LHCb collaboration),
Phys. Rev. Lett. \textbf{123} (2019) no.21, 211801

\end{thebibliography}
\end{document}